\documentclass[12pt]{article}
\usepackage{tikz}
\usepackage{amsmath,amssymb}
\usepackage{bbm}  
\usepackage{fancyhdr}
\usepackage[utf8]{inputenc}    
\usepackage{fullpage}
\usepackage{float}
\usepackage{authblk}
\usepackage{color}

\title{A distribution approach to finite-size corrections in Bethe Ansatz solvable models}
\author[1,2]{Etienne Granet}
\author[1,2,3]{Jesper Lykke Jacobsen}
\author[1,4]{Hubert Saleur}
\affil[1]{Institut de physique th\'eorique, Universit\'e Paris Saclay, CEA, CNRS, F-91191 Gif-sur-Yvette}
\affil[2]{Laboratoire de physique th\'eorique, D\'epartement de physique de l'ENS,
\'Ecole Normale Sup\'erieure, UPMC Univ.~Paris 06, CNRS, PSL Research University, 75005 Paris, France}
\affil[3]{Sorbonne Universit\'es, UPMC Univ.~Paris 06, \'Ecole Normale
Sup\'erieure, CNRS, Laboratoire de Physique Th\'eorique (LPT ENS), 75005 Paris, France}
\affil[4]{USC Physics Department, Los Angeles CA 90089, USA}
\date{}

\begin{document}
\maketitle
\begin{abstract}
We present a new and efficient method for deriving finite-size effects in statistical physics models solvable by Bethe Ansatz.
It is based on the study of the functional that maps a function to the sum of its evaluations over the Bethe roots.
A simple and powerful constraint is derived when applying this functional to infinitely derivable test functions
with compact support, that generalizes then to more general test functions.

The method is presented in the context of the simple spin-$1/2$ XXZ chain for which we derive the finite-size corrections 
to leading eigenvalues of the Hamiltonian for any configuration of Bethe numbers with real Bethe roots. The expected results for  the central charge and  conformal dimensions  are recovered.

\end{abstract}
\section{Introduction}
The use of field theory to study  long-distance properties of lattice models near criticality is one of the best established tools of statistical physics. 
It is possible to inverse the logic, i.e. to gain understanding of some field theories by studying well-chosen lattice models for which analytical computations are possible. This was illustrated, for instance, in a recent work where the black-hole sigma model of string theory was tackled using a special kind of spin-chain \cite{IJS}. A good deal of the current work on logarithmic conformal field theory (LCFT) stems from this idea.\\

The relationship between the lattice models and the field theory limit is particularly transparent in two ($1+1$) dimensions when conformal invariance is present. In this case, for large but finite size,  crucial information about the underlying CFT - like the central charge or the conformal dimensions \cite{BPZ} -  appears in the asymptotic expansion of certain physical quantities such as the (logarithms of) transfer matrix eigenvalues \cite{CardyCFTAmplitude}. For models solvable by Bethe Ansatz \cite{Bethe} these asymptotic expansions can be sometimes carried out analytically and thus reveal the structure of the field theory. Although thermodynamic properties are efficiently computed with Bethe root densities, the calculation of finite-size effects  remains a demanding and subtle task. This hampers progress on the understanding, for instance, of models having a non-compact continuum limit \cite{VJS}. \\

Two main methods have been developed so far.\\

The first (from a historical point of view) is the so-called ``Wiener-Hopf method'', developed in \cite{WoynaEckle, DevegaWoyna, Karowski, HamerQuispelBatchelor}. This method involves Bethe root densities, Euler-MacLaurin formula and integral Wiener-Hopf equations. Its main drawback is that it cannot be applied to cases where the Bethe roots are not real but arranged into complex conjugate pairs denoted as ``strings'', like for example in the spin $S>{1\over 2}$ Heisenberg model \cite{ReshKiri}. Moreover, as noticed in \cite{Karowski}, some terms which were originally neglected in the method  turned out to be non-negligible after all, affecting some of the intermediate results (such as the relation between the largest Bethe root and the density at this point, according to \cite{Karowski}), although they do not appear to have consequences on the final result.\\

The second method is the ``Non-Linear Integral Equation method'' (NLIE), developed in \cite{NLIEKlumperBatchelor,PearceKlumper,Klumper,Klumper2, DestriDevega}. Here, one exploits the analyticity properties of the eigenvalues of the transfer matrix in terms of the spectral parameter to derive an NLIE for the counting function. Besides the efficient numerical algorithms based on this method, it has been successfully applied to some higher-rank systems \cite{PZJ} and some  cases involving strings, such as the spin-$S$ Heisenberg model \cite{Klumper,Suzuki}. However there is no general recipe to derive NLIE equations for a new model, and, to our knowledge, very little is known on how to adapt the method to cases with isolated Bethe roots, or to the computation of  higher-order corrections.\\

We present in this paper  a new and efficient method for computing finite-size effects. Our approach is crucially based on the study of the functional that maps a function to the sum of its evaluation over the Bethe roots, seen as a {\sl distribution}. It involves moreover two crucial points. The first  is the observation that a simple and powerful constraint on this distribution can be derived by applying it to  infinitely differentiable functions with compact support (and then  to more general functions): it is the sum, with coefficients to be determined, of the Fourier transform of the functions at very particular values that depend on the Bethe Ansatz equations. The second point is to observe that this distribution evaluates very simply on the counting function itself, leading to an equation for these coefficients.\\

The method can be applied to higher-rank systems as soon as the Bethe roots are real. Possible adaptations of this method to the case of complex roots in some higher-rank or higher-spin Bethe equations \cite{Essler, PZJ} should be discussed elsewhere. It is our hope that this new method could eventually make possible analytical calculations e.g.\ of non-compact spectra and densities of states in models such as the one studied in \cite{VJS}.\\

For pedagogical reasons, we discuss all the relevant details in the case of the spin-$1/2$ XXZ spin chain with periodic or twisted boundary conditions, deriving the well-known central charge and conformal dimensions.

\section{Problem setting}

We consider the spin-$1/2$ XXZ spin chain on $L$ sites with twisted boundary conditions, and Hamiltonian 
\begin{equation}
\label{H}
H=-\dfrac{1}{2\sin\gamma}\sum_{k=1}^L\left(\sigma^x_k\sigma^x_{k+1}+\sigma^y_k\sigma^y_{k+1}+\Delta (\sigma^z_k\sigma^z_{k+1}-1) \right)\,,
\end{equation}
where the $\sigma^i_k$ are the Pauli matrices and $\Delta=-\cos\gamma$ an anisotropy parameter. The boundary conditions are
\begin{equation}
 \sigma_{L+1}^x\pm i\sigma_{L+1}^y=e^{-2i\pi\varphi}(\sigma_1^x\pm i\sigma_1^y)\quad \text{,}\quad \sigma^z_{L+1}=\sigma^z_1\,.
\end{equation}
This model is solvable by Bethe Ansatz \cite{Bethe,Baxter}: an eigenvalue $E_L$ of $H$ can be written as $E_L=Le_L$ with
\begin{equation}
e_L=-\dfrac{2\pi}{L}\sum_{i=1}^M s'(\lambda_i)\,,
\end{equation} 
where the function $s'$ will be given below, and the Bethe roots $\lambda_i$ are the $M$ solutions of the Bethe equations:
\begin{equation}
\left(\dfrac{\sinh(\lambda_i+i\gamma/2)}{\sinh(\lambda_i-i\gamma/2)} \right)^L=e^{-2i\pi\varphi}\prod_{j=1,j\neq i}^{M}\dfrac{\sinh(\lambda_i-\lambda_j+i\gamma)}{\sinh(\lambda_i-\lambda_j-i\gamma)}\,.
\end{equation}
Note that both $M$ and the $\lambda_i$'s implicitly depend on $L$. Taking the $\log$ of these equations transforms the products into sums, and leads to the  accumulation of  a multiple of $2\pi i$ because of $\log(zz')=\log z+\log z'+2in\pi$ with $n\in\{-1,0,1\}$. Thus the Bethe equations can be written as
\begin{equation}
\label{n0}
\begin{aligned}
 &z_L(\lambda_i)=\dfrac{I_i}{L}\text{ , }i=1,\ldots,M\\
 \text{where: }&z_L(\lambda)=s(\lambda)-\dfrac{1}{L}\sum_{i=1}^{M}r(\lambda-\lambda_i)+\dfrac{\varphi}{L}\,.
\end{aligned}
\end{equation}
$I_i$ are integers if $M$ is odd, half-integers if $M$ is even (because of the $-1$ in the definition of $r$ below), and are called Bethe numbers. $z_L$ will be called the {\sl counting function}. The functions $s$ and $r$ are

\begin{equation}
\begin{aligned}
s(\lambda)&=-\dfrac{1}{2i\pi}\log \left(-\dfrac{\sinh(\lambda+i\gamma/2)}{\sinh(\lambda-i\gamma/2)} \right)=\dfrac{1}{\pi}\arctan \left(\dfrac{\tanh \lambda}{\tan \gamma/2} \right) \,, \\
r(\lambda)&=-\dfrac{1}{2i\pi}\log \left(-\dfrac{\sinh(\lambda+i\gamma)}{\sinh(\lambda-i\gamma)} \right)=\dfrac{1}{\pi}\arctan \left(\dfrac{\tanh \lambda}{\tan \gamma} \right)\,.
\end{aligned}
\label{s-r-functions}
\end{equation}
The branch cut of the logarithm is taken such that $\log e^{i\theta}=i\theta $ for $-\pi<\theta\leq \pi$. Observe  that if we add $1/2$ to the twist $\varphi$ in $z_L$ when $M$ is odd then the Bethe numbers $I_k$ are half-integers. For this reason in the following we will always assume that the Bethe numbers are half-integers.

The relevant physical information about the CFT that describes the chain in the continuum limit is the term in $L^{-2}$ in the asymptotic expansion of $e_L$. To study this term, we define the following {\sl functional for a test function $\phi$}
\begin{equation}
S_L(\phi)=\dfrac{1}{L}\sum_{i=1}^M\phi(\lambda_i)\,,
\end{equation}
Noting that $e_L=-2\pi S_L(s')$, our objective is thus  to determine the asymptotic expansion of $S_L$ at large $L$. We will focus on the ground state and the first excited states, thus on states for which $M=L/2-n$ with finite $n$.

\subsection*{Notations}
We give here a few notations.\\
The Fourier transform $\hat{f}$ of a function $f$ is defined as
\begin{equation}
\hat{f}(\omega)=\int_{-\infty}^\infty f(x)e^{i\omega x}dx\,.
\end{equation}\\
Note that with this convention we have $\hat{f'}(\omega)=-i\omega \hat{f}(\omega)$. The convolution $f\star g$ of two functions is
\begin{equation}
(f\star g)(x)=\int_{-\infty}^\infty f(y)g(x-y)dy\,.
\end{equation}\\
For $y \in \mathbb{R}$, we denote by $f(y-\cdot)$ the $y$-dependent function of $x$ given by $x\to f(y-x)$.\\
For a function $\phi$ we introduce the notation
\begin{equation}
\phi_{\pm\infty}=\underset{x\to\pm\infty}{\lim}\phi(x)\quad \text{for a function }\phi\,,
\end{equation}
with the shorthand $\phi_\infty=\phi_{+\infty}$ when the function $\phi$ is odd (this should not be  confused  with $z_\infty$ and $S_\infty$ introduced in the following).

\subsection*{A few reminders about distributions}
We give here a brief reminder about distributions.\\
A distribution $T$ is a continuous linear map from the infinitely differentiable functions with compact support $\phi$ (called ``test functions''), to the complex numbers $\mathbb{C}$.\\
The most widely known example of a distribution is the Dirac delta (at $0$) $\delta_0$ that is defined by
\begin{equation}
\delta_0(\phi)=\phi(0)\,.
\end{equation}
Every locally integrable function $f$ also defines a distribution $T_f$ (or simply still denoted $f$) by
\begin{equation}
T_f(\phi)=\int^\infty_{-\infty} f(x)\phi(x)dx\,.
\end{equation}
The derivative $T'$ of a distribution $T$ is defined by
\begin{equation}
T'(\phi)=-T(\phi')\,.
\end{equation}
The convolution between a distribution $T$ and a test function $f$ is itself a function defined by
\begin{equation}
\label{convolution}
(T\star f)(y)=T(f(y-\cdot))\,.
\end{equation}
It can be seen as a distribution and with $\tilde{f}(x)=f(-x)$ can be rewritten
\begin{equation}
\label{convolution2}
(T\star f)(\phi)=\int^\infty_{-\infty} T(f(y-\cdot)) \phi(y)dy=T(\tilde{f}\star \phi)\,.
\end{equation}
If a distribution $T$ can be applied on rapidly decreasing functions $\phi$ (that satisfy $|x^n \phi^{(p)}(x)|\to 0$ as $x\to\pm\infty$ for all integers $n,p$) then its Fourier transform $\hat{T}$ is defined as
\begin{equation}
\hat{T}(\phi)=T(\hat{\phi})\,.
\end{equation}
An important property of distributions is that if $T$ satisfies
\begin{equation}
T\cdot x^n=0
\end{equation}
then there exist $n$ complex numbers $a_0,...,a_{n-1}$ such that
\begin{equation}
T=\sum_{i=0}^{n-1} a_i\delta_0^{(i)}
\end{equation}

\section{Leading-order term}
Define $S_\infty(\phi)$ by:
\begin{equation}
 S_\infty(\phi)=\underset{L\to\infty}{\lim}S_L(\phi)\,.
\end{equation}
The computation of $S_\infty(\phi)$ is carried out upon the assumption that $z_L$ converges to a function $z_\infty$ as $L\to\infty$, as is done to determine $e_\infty=\lim_{L\to\infty} e_L$ \cite{Lieb,Reshetikhin}. Then $\lambda_i\sim z_\infty^{-1}(I_i/L)$, so that:
\begin{equation}
 S_L(\phi)\underset{L\to\infty}{\sim} \dfrac{1}{L}\sum_{i=1}^{M}\phi(z_\infty^{-1}(I_i/L))\,.
\end{equation}
This is a Riemann sum that converges to $\int \phi\circ z_\infty^{-1}$, the integration limits being the limit of $I_1/L$ and $I_{M}/L$ as $L\to\infty$. A change of variable $\lambda=z_\infty^{-1}(x)$ gives $\int_{Q_-}^{Q_+} \phi z_\infty'$ with $Q_-=z_\infty^{-1}(I_1/L),Q_+=z_\infty^{-1}(I_M/L)$,
which motivates the definition of the density of roots:
\begin{equation}
\sigma_\infty=z_\infty'.
\end{equation}
Moreover for the ground state and first excited states we have $Q_\pm\to \pm\infty$ \cite{Lieb}, giving:
\begin{equation}
\label{n1}
 S_\infty(\phi)=\int_{-\infty}^{\infty}\phi\sigma_\infty\,.
\end{equation}
To compute $\sigma_\infty$, we use formula \eqref{n1} in equation \eqref{n0} for $L\to\infty$. This leads to:
\begin{equation}
 \sigma_\infty=s'-\sigma_\infty\star r'\,,
\end{equation}
which can be solved by Fourier transform:
\begin{equation}
\label{aer2}
 \hat{\sigma}_\infty=\dfrac{\widehat{s'}}{1+\widehat{r'}}\,.
\end{equation}
In this model a simple expression can be found:
\begin{equation}
\label{sigmaxxz}
\sigma_\infty(x)=\dfrac{1}{2\gamma\cosh(\pi x/\gamma)}\,.
\end{equation}
For notational convenience we introduce
\begin{equation}
\label{alpha}
\pm \alpha=\underset{x\to\pm\infty}{\lim}z_\infty(x)\,,
\end{equation}
and $v_F>0$ (the ``Fermi'' or ``sound'' velocity, as will be explained further) the constant such that
\begin{equation}
\label{defvf}
\sigma_\infty(x)\propto e^{-v_F |x|}\quad \text{as }x\to \pm\infty\,.
\end{equation}
According to \eqref{sigmaxxz}, we have then $v_F = \pi / \gamma$. For the spin-$1/2$ XXZ model we have $\alpha=1/4$.

We now would like to compute the next-to-leading-order term in $S_L(\phi)$. To that end we need to know the finite-size corrections of a Riemann sum $\sum_k f(k/L)$. Before going any further, we recall some results on this question.\\

\section{Reminders about  Riemann sums \label{riemann}}
\subsection*{Euler-Maclaurin formula}
For $f$ a function defined on $[0,1]$ and $t\in [0,1]$, we define the $t$-shifted Riemann sum
\begin{equation}
R_L^t(f)=\dfrac{1}{L}\sum_{k=0}^{L-1}f\left(\dfrac{k+t}{L} \right)\,.
\end{equation}
If $f$ is $\mathcal{C}^{n}$ on $[0,1]$, then the Euler-MacLaurin formula reads as follows \cite{EulerMacLaurin}
\begin{equation}
R_L^t(f)=EM_L^{t}(f)+o(L^{-n})\,,
\end{equation}
with
\begin{equation}
EM_L^t(f)=\int_0^1f+\sum_{k=1}^{n}B_{k}(t)\dfrac{f^{(k-1)}(1)-f^{(k-1)}(0)}{k!L^{k}}\,,
\end{equation}
where $B_{k}(t)$ are the Bernoulli polynomials with exponential generating function
\begin{equation}
 \sum_{k=0}^\infty B_k(t) \frac{x^k}{n!} = \frac{x {\rm e}^{x t}}{{\rm e}^x - 1} \,.
\end{equation}
For example, up to order $O(L^{-2})$ it reads:
\begin{equation}
\label{eulermaclaurin2}
R_L^t(f)=\int_0^1 f+\dfrac{t-1/2}{L}(f(1)-f(0))+\dfrac{t^2/2-t/2+1/12}{L^2}(f'(1)-f'(0))+o(L^{-2})\,.
\end{equation}

\subsection*{Dependence on $L$}
In the previous formula $f$ is a function that does not depend on $L$, unlike in  our problem. Assume  now that we have the following expansion for a sequence of functions $f_L$
\begin{equation}
f_L(x)=g_0(x)+g_1(x)L^{-1}+ \cdots +g_n(x)L^{-n}+o(L^{-n})\,,
\end{equation}
with $g_i$ being $\mathcal{C}^{n-i}$ functions that do not depend on $L$. Assume that the convergence of this expansion is uniform, meaning that the remainder $\epsilon_L$
\begin{equation}
 \epsilon_L(x) = f_L(x) - \sum_{p=0}^n g_p(x) L^{-p}\,,
\end{equation}
is $o(L^{-n})$ and that $L^n |\epsilon_L(x)|$ can be bounded by a sequence $a_L\to 0$ uniformly in $x$ (otherwise it only implies that it goes to zero for a fixed $x$). Then $R^t_L(\epsilon_L)=o(L^{-n})$. It follows that
\begin{equation}
R^t_L(f_L)=EM^t_L(g_0)+EM^t_L(g_1)L^{-1}+ \cdots +EM^t_L(g_n)L^{-n}+o(L^{-n})\,,
\end{equation}
or in a more compact form
\begin{equation}
\label{eml}
R^t_L(f_L)=EM_L^t(f_L)+o(L^{-n})\,,
\end{equation}
meaning that uniform convergence allows us to use the Euler-MacLaurin formula for $L$-dependent functions.

\subsection*{Uniform convergence}
Unfortunately we do not have uniform convergence in Bethe equations. Indeed we observe numerically that for fixed $x$ we have
\begin{equation}
z_L(x)=z_\infty(x)+\dfrac{g(x)}{L^{\beta}}+o(L^{-\beta})\,,
\end{equation}
with $\beta>1$ and some function $g$, whereas $\max_{x} |z_L(x)-z_\infty(x)|$ decreases more slowly than $L^{-\beta}$.
The consequence for the Bethe roots is that, although most of them are separated by $O(L^{-1})$, this is not the case for the extremal ones: a numerical computation of the Bethe roots shows that their separation is greater than $L^{-1}$.\\

But Dini's theorem saves us partly \cite{Dini}: a sequence of continuous and increasing functions that converges simply to a continuous function on a segment $[a,b]$ also converges uniformly. Since $z_L$ converges to an increasing function $z_\infty$, on a fixed segment $[a,b]$ it is increasing for $L$ large enough and thus $z_L$ converges uniformly on any segment. This  means that Euler-MacLaurin formula could be applied if we were to sum only over Bethe roots smaller in absolute value than some fixed $\lambda$. This suggests to first look at $S_L(\phi)$ for $\phi$ having compact support on $\mathbb{R}$. Since the functions to which we want to apply $S_L$ do not have this property of compactness, we will have to extend  the result  to more general functions in several steps: first to test functions with possible discontinuities, second to functions without compact support but that decrease exponentially fast at infinity, third to functions that converge to a non-zero value exponentially fast at infinity.

\section{Constraint on the next-to-leading-order}
\subsection{$\mathcal{C}^\infty$ test functions with compact support}
Define $w_L(\phi)$, the distribution to be studied, as:
\begin{equation}
 w_L(\phi)=S_L(\phi)-S_\infty(\phi) \,.
\end{equation}
Let $\phi$ be a $\mathcal{C}^\infty$ function with compact support. Summing over all $\lambda_i$'s is the same as summing over the $\lambda_i$'s in the support of $\phi$. But $z_L$ is invertible on this segment, so that we can write $\lambda_i=z_L^{-1}(I_i/L)$. Then using eq.~\eqref{eml}
\begin{equation}
 S_L(\phi)=EM_L^{1/2}(\phi\circ z_L^{-1})+o(L^{-n})\,,
\end{equation}
for every $n$, since $\phi$ and $z_L$ are $\mathcal{C}^\infty$ functions. And since $\phi$ has compact support we have
\begin{equation}
S_L(\phi)=\int^\alpha_{-\alpha} \phi\circ z_L^{-1}\,,
\end{equation}
which becomes, after a change of variables,
\begin{equation}
\label{n3}
 S_L(\phi)=\int_{-\infty}^{+\infty} \phi z_L'\,.
\end{equation}
But $z_L'$ can be obtained through eq.~\eqref{n0}:
\begin{equation}
\label{n4}
 z_L'(\lambda)= \sigma_\infty(\lambda)-w_L (r'(\lambda-\cdot))\,. 
\end{equation}
This gives the equation valid for all $\mathcal{C}^\infty$ functions $\phi$ with compact support
\begin{equation}
(w_L+w_L\star r')(\phi)=0\,,
\label{rel-with-compact-supp}
\end{equation}
where the convolution $w_L\star r'$ is defined in eq.~\eqref{convolution} and eq~\eqref{convolution2}.
For the Fourier transform, this implies that, as a distribution,
\begin{equation}
 \widehat{w}_L(1+\widehat{r'})=0\,,
\end{equation}
so that:
\begin{equation}
\label{gen12}
\widehat{w}_L=\sum_{\omega\in\Omega}A_{\omega}\delta_{\omega}\,,
\end{equation}
where
\begin{equation}
\label{omega}
\Omega=\{ \omega\in\mathbb{C},1+\hat{r'}(\omega)=0 \}\,,
\end{equation} 
and $\delta_\omega$ is the Dirac distribution at $\omega$. The $A_{\omega}$ denotes a set of ``constants'' to be determined. They are constants in the sense that they do not depend on the function $\phi$ to which $w_L$ is applied, but they clearly have to depend on $L$. In the XXZ model $\Omega$ can be computed to be
\begin{equation}
\Omega=\left\lbrace \dfrac{2ni}{1-\gamma/\pi}, n\in\mathbb{Z} \right\rbrace \cup \left\lbrace \dfrac{\pi}{\gamma}(2m+1)i, m\in\mathbb{Z} \right\rbrace \,.
\end{equation} 
It is assumed in eq. \eqref{gen12} for notational convenience that $1+\hat{r'}$ has only simple poles---otherwise derivatives of $\delta$ would be present---which is true as soon as $\gamma/\pi$ is irrational (the case with derivatives is treated at the end of section \ref{doublezero}). Eq.~\eqref{gen12} is true up to an `exponentially small' correction that is negligible compared to  any power $L^{-n}$. Explicitly, eq.~\eqref{gen12} means that for $\mathcal{C}^\infty$ functions with compact support
\begin{equation}
S_L(\phi)=S_\infty(\phi)+\sum_{\omega\in\Omega}A_\omega \hat{\phi}(\omega)\,.
\end{equation}

\subsection{Discontinuous test functions with compact support}
To compute the $A_{\omega}$ we need some `boundary condition': by this we mean a function $f$ for which we know $ w_L(f)$. It is almost the case of $z_L$, since $z_L(\lambda_i^L)=I_i/L$ exactly (note: this  is not an approximation in $L$). But in order to gain more information we would like to sum only over Bethe roots with positive or negative Bethe numbers. And this leads to applying the distribution $w_L$ to discontinuous functions (at a finite number of points), which is outside the standard theory of distributions.\\

The standard theory of distributions indeed applies to $\mathcal{C}^\infty$ test functions. Whenever discontinuities occur, some distributions such as the Dirac delta become ill-defined, and additional terms come from integration by parts. We use here partly the work of \cite{Kurasov}. Without loss of generality, we assume---like in \cite{Kurasov}---that the test functions have a discontinuity at $0$ (otherwise one just has to shift the function). The two main differences with the usual distributions are the Dirac distribution, which is now defined as:
 \begin{equation}
  \delta_0(\phi)=\dfrac{\phi(0+)+\phi(0-)}{2}\,,
 \end{equation}
where $\phi(0\pm)$ is the right or left limit of $\phi$ at $0$, and the distribution $\beta$ defined as:
 \begin{equation}
  \beta(\phi)=\phi(0+)-\phi(0-)\,.
 \end{equation}
 It satisfies $\beta=1'$, the derivative of the constant distribution. The Fourier transform of $\beta$ will be needed. It is defined as usual by $\hat{\beta}(\phi)=\beta(\hat{\phi})$. Thus $\hat{\beta}(\phi)$ is proportional to the coefficient in front of $1/x$ in the expansion of $\phi(x)$ when $x\to\infty$, as can be seen from the fact that the Fourier transform of the Heaviside function is $-(i\omega)^{-1}$.

Let now $\phi$ be a function with compact support and a discontinuity at zero. Let us look at the value of the Bethe roots close to zero. We have thus $\lambda_i=z_L^{-1}(I_i/L)$ with $I_i/L\to 0$, which allows for a Taylor expansion. Using $z_L(z_L^{-1}(0))=0$ we get
\begin{equation}
\label{approx}
\lambda_i=z_L^{-1}(I_i/L)=\dfrac{I_i/L-z_L(0)}{\sigma_\infty(0)}+o(L^{-1})\,.
\end{equation}
We see here that $z_L(0)$ acts like a shift for the Bethe numbers, and thus deserves a specific notation. Define $\varphi_0$ through
\begin{equation}
z_L(0)=\dfrac{\varphi_0}{L}+o(L^{-1})\,,
\end{equation}
where $\varphi_0$ can take the value $0$ if $z_L(0)$ decreases faster than $L^{-1}$. Recall now that the Bethe numbers are assumed to be half-integers. Thus the sum over the Bethe roots around zero is a $t$-shifted Riemann sum with $t=1/2-\varphi_0$. Applying eq.~\eqref{eml} with such $t$ on $[K_-,0]$ and on $[0,K_+]$, where the support of $\phi$ is $[K_-,K_+]$, gives at order $o(L^{-2})$
\begin{equation}
S_L(\phi)=\int \phi z_L'-B_1(t)\dfrac{\phi(0^+)-\phi(0^-)}{L}-B_2(t)\dfrac{\phi'(0^+)-\phi'(0^-)}{2\sigma_\infty(0)L^2}+o(L^{-2})\,.
\end{equation}
This implies that (\ref{rel-with-compact-supp}) has the leading finite-size correction
\begin{equation}
(w_L+w_L\star r')(\phi)=-\dfrac{B_1(t)\beta(\phi)}{L}-\dfrac{B_2(t)\beta(\phi')}{2\sigma_\infty(0)L^2}+o(L^{-2})\,.
\end{equation}
Upon Fourier transforming, since $1+\hat{r'}(\omega)\to 1$ when $\omega\to\infty$, we have $\hat{\beta}(1+\hat{r'})^{-1}=\hat{\beta}$, yielding:
\begin{equation}
\label{gen12discon1}
w_L(\phi)=\sum_{\omega\in\Omega}A_\omega \hat{\phi}(\omega)-\dfrac{B_1(t)\beta(\phi)}{L}-\dfrac{B_2(t)\beta(\phi')}{2\sigma_\infty(0)L^2}+o(L^{-2})\,.
\end{equation}
In the following we will encounter functions $\phi$ with discontinuity at $t_0/L$ with $t_0$ real. We recover case \eqref{gen12discon1} by applying it to $\phi(\cdot-t_0/L)$, which amounts to setting $t=1/2-\varphi_0-t_0\sigma_\infty(0)$. In particular if the discontinuity is at $z_L^{-1}(0)=-z_L(0)/\sigma_\infty(0)+o(z_L(0))$ then we have $t=1/2$. This case will be particularly useful and we get
\begin{equation}
\label{gen12discon}
w_L(\phi)=\sum_{\omega\in\Omega}A_\omega \hat{\phi}(\omega)+\dfrac{\beta(\phi')}{24\sigma_\infty(0)L^2}+o(L^{-2})\,.
\end{equation}

\subsection{Test functions with exponential decay \label{expdecay}}
We would like now to extend the application of $w_L$ to $\mathcal{C}^\infty$ functions $\phi$ without compact support that behave as $\phi(x)\sim ce^{-a |x|}$ with $c$ and $a>0$ some constants, as $x\to\pm\infty$.\\

Let us assume in a first step that the function $\phi$ is $\mathcal{C}^\infty$ without compact support, but such that it satisfies $\phi(x)=o(e^{-a |x|})$ for all $a>0$, meaning that it decreases faster than any exponential at infinity. Notice that $\hat{\phi}$ is defined in the complex plane without singularities in this case, and that conversely if $\hat{\phi}$ is defined in the complex plane without singularities then $\phi$ has to decay faster than any exponential. For every $K>0$ define a test function $\phi_K$ $\mathcal{C}^\infty$ with compact support such that $\phi_K(x)=\phi(x)$ for the reals $|x|<K$ and $\phi_K(x)=0$ for $|x|>K+\epsilon$, where $\epsilon$ is a small parameter. For a real $a$ we have $\hat{\phi}_K(ia)=\int e^{-ax}\phi_K(x)dx \to \int e^{-ax}\phi(x)dx=\hat{\phi}(ia)$ when $K\to\infty$, because $\phi$ decreases faster than any exponential (otherwise these integrals would diverge for $a$ large enough and analytic continuation has to be done). Then formula \eqref{gen12} applied to $\phi_K$ gives a more and more accurate approximation of $w_L(\phi)$ as $K\to\infty$, and converges to the same formula applied to $\phi$. Thus eq.\eqref{gen12} is still valid for functions decreasing faster than any exponential. \\

Now consider a function $\phi$ with exponential decay: $\phi(x)\sim \sum_n c_ne^{-xv_n}$ with $c_n$ and $v_n>0$ some constants, when $x\to\infty$. Then by adding and subtracting functions $e^{-a x}\mathbbm{1}_{x\geq 0}$, whose Fourier transform is $(-i\omega+a)^{-1}$, we see that $\hat{\phi}$ is a function in the complex plane having poles at $-iv_n$ with residues $c_n$. Let now $f$ be an even function such that $\hat{f}$ is analytic and $\hat{f}(-iv_n)=0$. The function $\phi\star f$ decays faster than any exponential at infinity whenever $\phi$ has the previous decay properties, since its Fourier transform has no poles in the complex plane. We can then  apply formula \eqref{n3} to it:
\begin{equation}
S_L(\phi\star f)=\int (\phi\star f)z_L'\,.
\end{equation} 
Using the same arguments as previously, and $w_L(\phi\star f)=(w_L\star f)(\phi)$, we get:
\begin{equation}
\hat{w_L}\hat{f}(1+\hat{r'})=0\,,
\end{equation}
implying that the distribution $\hat{w_L}$ has its support in the set of zeros of $\hat{f}(1+\hat{r'})$. Since $\hat{\phi}$ has a pole at the zeros of $\hat{f}$, we have
\begin{equation}
\label{gen12decay}
w_L(\phi)=\sum_{\omega\in\Omega}A_\omega \hat{\phi}(\omega)+\sum_{\omega\in\Omega_\phi}A_\omega \text{Res}_\omega(\hat{\phi})\,,
\end{equation}
with $\Omega$ the set of zeros of $1+\hat{r'}$, and $\Omega_\phi$ the set of poles of $\hat{\phi}(1+\hat{r'})^{-1}$ without the set of zeros of $(1+\hat{r'})$. $\text{Res}_\omega(\hat{\phi})$ is the residue of $\hat{\phi}$ at $\omega$.

Notice than whenever $\hat{\phi}$ has the same poles as $1+\hat{r'}$ (which is the case for $\hat{s'}$ and $\hat{r'}$, see eq.~\eqref{s-r-functions}) there is no additional terms compared to \eqref{gen12}. It is also assumed for simplicity of notation that there are no multiple poles, otherwise additional derivatives of Dirac deltas must be considered.\\

In the event of discontinuities, the function $\phi$ can be decomposed as $\phi=\phi_r+\phi_c$ with $\phi_r$ without discontinuities (but without compact support) and $\phi_c$ with compact support (but with discontinuities), enabling  one to combine equations \eqref{gen12decay} and \eqref{gen12discon}.\\

Let us now derive an important bound on the $A_\omega$. Indeed remark that since $z_L'(\lambda)=\sigma_\infty(\lambda)-w_L(r'(\lambda-\cdot))$ we can express $z_L'$ in terms of the $A_\omega$. Since $\hat{r'}$ has the same poles as $1+\hat{r'}$, the poles of $\hat{r'}/(1+\hat{r'})$ are the zeros of $1+\hat{r'}$. Since by definition $\hat{r'}(\omega)=-1$ for $\omega\in\Omega$, we get
\begin{equation}
z_L(\lambda)=z_\infty(\lambda)+\sum_{\omega\in\Omega}\dfrac{e^{i\omega\lambda}}{i\omega}A_\omega\,.
\end{equation}
Observe now that for $\Lambda_L$ the largest root we have at leading order $z_\infty(\Lambda_L)\sim I_M/L$. This implies
\begin{equation}
\label{lambdaL}
\Lambda_L=\dfrac{\log L}{v_F}+o(\log L)\,.
\end{equation}
Since $z_L(\Lambda_L)=I_M/L$ we deduce that $z_L(\Lambda_L)-z_\infty(\Lambda_L)$ is of order $O(L^{-1})$ and that for all $\omega\in\Omega$ we should have the bound $|A_\omega e^{i\omega\Lambda_L}|=O(L^{-1})$. With a similar analysis for the largest (in absolute value) negative root, and using eq.~\eqref{lambdaL} we get then the important bound
\begin{equation}
\label{bound}
A_\omega=O(L^{-1-|\omega|/v_F})\,.
\end{equation}
Notice  that to have the behaviour \eqref{defvf}, $\pm iv_F$ have to be elements of $\Omega$, and from eq.~\eqref{aer2} they even satisfy (with $\Im$ denoting the imaginary part):
\begin{equation}
v_F=\min \{\Im \omega, \quad \text{with } \omega \in \Omega\,, \Im\omega>0\,, \hat{s'}(\omega)\neq 0 \}\,.
\end{equation}

\subsection{Non-zero test functions at infinity \label{nonzero}}
The last but important extension of the formula \eqref{gen12decay} concerns functions $\phi$ that converge exponentially fast to non-zero values $\phi_{\pm\infty}$ at $\pm\infty$. In this case even the Fourier transform of $\phi$ becomes ill-defined.  There is a non-sophisticated way to get around the problem here. First note that since $z_L$ is increasing, a Bethe root $\lambda_i$ has a positive Bethe number $I_i$ if and only if $\lambda_i>z_L^{-1}(0)$. Let us now write $\phi$ as
\begin{equation}
\label{decompphi}
\phi=\phi_0+\phi_{+\infty}\mathbbm{1}_{>z_L^{-1}(0)}+\phi_{-\infty}\mathbbm{1}_{<z_L^{-1}(0)}\,,
\end{equation}
so that $\phi_0$ decays to zero at infinity, but has an additional discontinuity of $\phi_{-\infty}-\phi_{+\infty}$ at $z_L^{-1}(0)$. Then $w_L(\phi_0)$ is well-defined, and we only have to make sense of $w_L(\mathbbm{1}_{>z_L^{-1}(0)})$ and similarly of $w_L(\mathbbm{1}_{<z_L^{-1}(0)})$. By definition:
\begin{equation}
w_L(\mathbbm{1}_{>z_L^{-1}(0)})=S_L(\mathbbm{1}_{>z_L^{-1}(0)})-S_\infty(\mathbbm{1}_{>z_L^{-1}(0)})\,.
\end{equation}
Here $S_\infty(\mathbbm{1}_{>z_L^{-1}(0)})=\alpha-z_\infty(z_L^{-1}(0))$ and $S_L(\mathbbm{1}_{>z_L^{-1}(0)})= \frac{1}{L} \text{card}\{k,I_k>0\}$, i.e., the number of positive Bethe numbers, divided by $L$. Moreover using $z_L(z_L^{-1}(0))=0$ we have the expansion
\begin{equation}
z_L^{-1}(0)=-\dfrac{z_L(0)}{\sigma_\infty(0)}+o(z_L(0))\,.
\end{equation}
 Let us introduce $n_\pm$, the number of vacancies in positive or negative Bethe numbers, that is
\begin{equation}
\label{nnn}
\begin{aligned}
\text{card}\{I_k>0 \}&=L\alpha-n_+\\
\text{card}\{I_k<0 \}&=L\alpha-n_-\,.
\end{aligned}
\end{equation} 
We get:
\begin{equation}
\label{w1}
\begin{aligned}
w_L(\mathbbm{1}_{>z_L^{-1}(0)})&=-\left(\dfrac{n_+}{L}+z_L(0) \right)+o(z_L(0))\\
w_L(\mathbbm{1}_{<z_L^{-1}(0)})&=-\left(\dfrac{n_-}{L}-z_L(0) \right)+o(z_L(0))\,.
\end{aligned}
\end{equation}
But from eq.~\eqref{n0} we have:
\begin{equation}
z_L(0)=w_L(r)+\dfrac{\varphi}{L}\,,
\end{equation}
which gives, decomposing $r$ as in eq.\eqref{decompphi}, and applying eq.~\eqref{w1},
\begin{equation}
z_L(0)=w_L(r_0)-r_{\infty}\left(\dfrac{n_+}{L}+z_L(0) \right)+r_{\infty}\left(\dfrac{n_-}{L}-z_L(0)\right)+\dfrac{\varphi}{L} \,.
\end{equation}
Since the discontinuity of $r_0$ is at $z_L^{-1}(0)$, from eq.~\eqref{gen12discon} and the bound \eqref{bound}, we get that $w_L(r_0)$ is $o(L^{-1})$. It follows that at leading order $L^{-1}$ $z_L(0)$ can be deduced as
\begin{equation}
\label{zl0}
z_L(0)=\dfrac{\varphi-r_\infty(n_+-n_-)}{1+2r_\infty}\dfrac{1}{L}\,.
\end{equation}
Now apply the same procedure to a general function $\phi$: decompose it as in eq.~\eqref{decompphi} and use eq.~\eqref{w1} to obtain
\begin{equation}
\label{useful}
w_L (\phi)= w_L(\phi_0)-\phi_{+\infty}\left(\dfrac{n_+}{L}+z_L(0) \right)-\phi_{-\infty}\left(\dfrac{n_-}{L}-z_L(0) \right)\,,
\end{equation}
where $z_L(0)$ is explicitly given by eq.~\eqref{zl0}.

\section{Computing the constants $A_{\pm iv_F}$ \label{constantsAomega}}
We have now the tools to compute the main constants $A_\omega$ needed in eq.~\eqref{gen12}. The idea is to compute $S_L(z_L\mathbbm{1}_{>z_L^{-1}(0)})$ in two different ways: the first one uses $z_L(\lambda_i)=I_i/L$ exactly, while the second one uses $z_L(\lambda)=s(\lambda)-S_L(r(\lambda-\cdot))+\varphi/L$.

\subsection*{First way}
The quantity $S_L(z_L\mathbbm{1}_{>z_L^{-1}(0)})$ is simply the sum of positive Bethe numbers divided by $L^2$. Using the following summation formula
\begin{equation}
\sum_{k=0}^{m-1}\left( k+\dfrac{1}{2}\right)=\dfrac{m^2}{2}\,,
\end{equation}
we get
\begin{equation}
\label{concl1}
S_L(z_L\mathbbm{1}_{>z_L^{-1}(0)})=\dfrac{\alpha^2}{2}+\dfrac{n^{2}_+}{2L^2}-\dfrac{\alpha n_+}{L}+\dfrac{\Delta_+ I}{L^2}\,,
\end{equation}
with 
\begin{equation}
\label{iii}
\Delta_+ I=\sum_{I_k\geq 0} I_k-\sum_{J_k\geq 0} J_k\,,
\end{equation}
 where $J_k$ are the Bethe numbers when all the vacancies are at the outmost  positions (equivalently: there is no vacancy between two Bethe numbers; or $\sum_{J_k\geq 0} J_k$ is minimal for a fixed number of Bethe roots). $\Delta_+ I$ is also the number of times each vacancy has been moved one step away from the Fermi surface. For example $\Delta_+ I = 0$ for the ground state. \\

\subsection*{Second way}
Here we use $z_L(\lambda)=z_\infty(\lambda)-w_L(r(\lambda-\cdot))+\varphi/L$ and $S_L=S_\infty+w_L$ to decompose:
\begin{equation}
\label{decompose}
\begin{aligned}
S_L(z_L \mathbbm{1}_{>z_L^{-1}(0)})=& S_\infty(z_\infty \mathbbm{1}_{>z_L^{-1}(0)})\\
&-S_\infty(w_L\star r \mathbbm{1}_{>z_L^{-1}(0)})+w_L(z_\infty \mathbbm{1}_{>z_L^{-1}(0)})\\
&-w_L(w_L\star r \mathbbm{1}_{>z_L^{-1}(0)})\\
&+S_L(\varphi \mathbbm{1}_{>z_L^{-1}(0)})L^{-1}\,.
\end{aligned}
\end{equation}
The first term in this equation can be directly computed
\begin{equation}
S_\infty(z_\infty \mathbbm{1}_{>z_L^{-1}(0)})=\dfrac{\alpha^2}{2}-\dfrac{z_L(0)^2}{2}\,.
\end{equation}
As for the second term in eq.~\eqref{decompose}, we do the following interchange of order of summation (recall that $w_L$ can be expressed as finite sums and integral over $\sigma_\infty$)
\begin{equation}
\label{detail}
S_\infty((w_L\star r) \mathbbm{1}_{>z_L^{-1}(0)})=-w_L((\sigma_\infty \mathbbm{1}_{>z_L^{-1}(0)})\star r)\,,
\end{equation}
where the minus sign comes from the oddness of $r$. To be clear and get used to the notations, let us detail eq.~\eqref{detail}:
\begin{equation}
\begin{aligned}
S_\infty((w_L\star r) \mathbbm{1}_{>z_L^{-1}(0)})&=\int_{-\infty}^{\infty} \sigma_\infty(x)(w_L \star r)(x)\mathbbm{1}_{>z_L^{-1}(0)}(x)dx\\
&=\int_{-\infty}^{\infty} \sigma_\infty(x)w_L(r(x-\cdot))\mathbbm{1}_{>z_L^{-1}(0)}(x)dx\\
&=w_L\left( \int_{-\infty}^{\infty} \sigma_\infty(x)\mathbbm{1}_{>z_L^{-1}(0)}(x) r(x-\cdot)dx\right)\\
&=-w_L\left(\int_{-\infty}^{\infty} \sigma_\infty(x)\mathbbm{1}_{>z_L^{-1}(0)}(x) r(\cdot-x)dx\right)\\
&=-w_L( (\sigma_\infty\mathbbm{1}_{>z_L^{-1}(0)}) \star r )\,.
\end{aligned}
\end{equation}
An integration by part also gives
\begin{equation}
\label{eq78}
(\sigma_\infty \mathbbm{1}_{>z_L^{-1}(0)})\star r=(\alpha+z_L(0)) r(\cdot-z_L^{-1}(0))+((z_\infty-\alpha)\mathbbm{1}_{>z_L^{-1}(0)})\star r'\,.
\end{equation}
Using eq.~\eqref{w1}, the third term in eq.~\eqref{decompose} is expressed as
\begin{equation}
w_L(z_\infty \mathbbm{1}_{>z_L^{-1}(0)})=w_L((z_\infty-\alpha) \mathbbm{1}_{>z_L^{-1}(0)})-\alpha \left(\dfrac{n_+}{L}+z_L(0) \right)\,.
\end{equation}
The fifth term of eq.~\eqref{decompose} is computed with the definition of $n_\pm$ in eq.\eqref{nnn}
\begin{equation}
S_L(\varphi \mathbbm{1}_{>z_L^{-1}(0)})L^{-1}=\left(\dfrac{\alpha}{L}-\dfrac{n_+}{L^2} \right)\varphi\,.
\end{equation}
Gathering everything, and using $w_L(r)=z_L(0)-\varphi/L$ we get
\begin{equation}
\begin{aligned}
S_L(z_L \mathbbm{1}_{>z_L^{-1}(0)})=&\dfrac{\alpha^2}{2}+\dfrac{z_L(0)^2}{2}-\dfrac{\alpha n_+}{L}-\dfrac{\varphi}{L}\left(\dfrac{n_+}{L}+z_L(0) \right)+w_L\left[(\delta_0+r')\star ((z_\infty-\alpha)\mathbbm{1}_{>z_L^{-1}(0)})\right]\\
&-w_L((w_L\star r) \mathbbm{1}_{>z_L^{-1}(0)})\,.
\end{aligned}
\end{equation}
We now just have to compute the Fourier transform of $f$ defined by
\begin{equation}
f=(\delta_0+r')\star ((z_\infty-\alpha)\mathbbm{1}_{>z_L^{-1}(0)})\,.
\end{equation}
It has a discontinuity of $-\alpha$ at $z_L^{-1}(0)$, and a discontinuity of the derivative of $\sigma_\infty(0)$. Moreover the Fourier transform of its derivative is:
\begin{equation}
\label{fcomplicated}
\hat{f'}(\omega)=(1+\hat{r'})(\widehat{\sigma_\infty \mathbbm{1}_{>z_L^{-1}(0)}})(\omega)\,.
\end{equation}
Recall from eq.~\eqref{omega} that, by definition, the $\omega\in\Omega$ satisfy $1+\hat{r'}(\omega)=0$. Thus in $w_L(f)$ we recover the residues of the poles of $\widehat{\sigma_\infty \mathbbm{1}_{>z_L^{-1}(0)}}$. But the poles of this function are the poles in the negative half-plane of $\hat{\sigma_\infty}=\hat{s'}(1+\hat{r'})^{-1}$. Thus:
\begin{equation}
w_L\left[(\delta_0+r')\star ((z_\infty-\alpha)\mathbbm{1}_{>z_L^{-1}(0)})\right]=\dfrac{1}{24 L^2}-\sum_{\omega\in\Omega_-}\dfrac{\hat{s'}(\omega)}{i\omega}A_\omega\,,
\end{equation} 
where we used the following notation 
\begin{equation}
\Omega_-=\{ \omega \in \mathbb{C}, 1+\hat{r'}(\omega)=0\quad\text{and }\Im\omega<0 \}\,.
\end{equation}
As for the term $w_L((w_L\star r) \mathbbm{1}_{>z_L^{-1}(0)})$ in eq.~\eqref{decompose}, the oddness of $r$ allows us to rewrite it as:
\begin{equation}
w_L((w_L\star r) \mathbbm{1}_{>z_L^{-1}(0)})=w_L\left[ w_L(r(\cdot-\cdot)\mathbbm{1}_{<z_L^{-1}(0)})\mathbbm{1}_{>z_L^{-1}(0)}\right]\,,
\end{equation}
where the first $w_L$ applies on the first dot $\cdot$. It means that the $w_L$ on the left applies on the function $x\to w_L(r(x-\cdot)\mathbbm{1}_{<z_L^{-1}(0)})\mathbbm{1}_{x>z_L^{-1}(0)}$. Since it involves two $w_L$, the only  non-zero term at order $O(L^{-2})$ is given by the $L^{-1}$ term in \eqref{useful}. Equation \eqref{useful} gives:
\begin{equation}
w_L(r(x-\cdot)\mathbbm{1}_{<z_L^{-1}(0)}))=-r_\infty \left(\dfrac{n_-}{L}-z_L(0) \right)\,,
\end{equation}
and then the second application of $w_L$ gives:
\begin{equation}
w_L((w_L\star r) \mathbbm{1}_{>z_L^{-1}(0)})=r_\infty \left(\dfrac{n_-}{L}-z_L(0) \right)\left(\dfrac{n_+}{L}+z_L(0) \right)\,.
\end{equation}
Gathering everything we obtain
\begin{equation}
\label{concl2}
\begin{aligned}
S_L(z_L \mathbbm{1}_{>z_L^{-1}(0)})=&\dfrac{\alpha^2}{2}+\dfrac{z_L(0)^2}{2}-\dfrac{\alpha n_+}{L}+\dfrac{1}{24 L^2}-\dfrac{\varphi}{L}\left(\dfrac{n_+}{L}+z_L(0) \right)\\
&-\sum_{\omega\in\Omega_-}\dfrac{\hat{s'}(\omega)}{i\omega}A_\omega-r_\infty \left(\dfrac{n_-}{L}-z_L(0) \right)\left(\dfrac{n_+}{L}+z_L(0) \right)\,.\\
\end{aligned}
\end{equation}

\subsection*{Conclusion}
Equating expressions \eqref{concl1} and \eqref{concl2} gives an equation for the $A_\omega$. Note that by definition of $v_F$, we have $\hat{s'}(\omega)=0$ if $|\omega|<v_F$. Moreover the bound \eqref{bound} ensures that the $A_\omega$ with $|\omega|>v_F$ do not appear in the expansion of the energy $e_L$ at order $L^{-2}$. Thus $A_{-iv_F}$ can be readily deduced
\begin{equation}
\label{preresult}
\begin{aligned}
\dfrac{\hat{s'}(-iv_F)A_{-iv_F}}{-v_F}=&-\dfrac{1}{24 L^2}+\dfrac{n^{2}_+}{2 L^2}+\dfrac{\varphi}{L}\left(\dfrac{n_+}{L}+z_L(0) \right)\\
&+r_\infty \left(\dfrac{n_-}{L}-z_L(0) \right)\left(\dfrac{n_+}{L}+z_L(0) \right)-\dfrac{1}{2}z_L(0)^2+\dfrac{\Delta_+ I}{L^2}\,,
\end{aligned}
\end{equation}
while a similar computation for $S_L(z_L \mathbbm{1}_{<z_L^{-1}(0)})$ gives
\begin{equation}
\begin{aligned}
\dfrac{\hat{s'}(iv_F)A_{iv_F}}{-v_F}=&-\dfrac{1}{24 L^2}+\dfrac{n^{2}_-}{2 L^2}-\dfrac{\varphi}{L}\left(\dfrac{n_-}{L}-z_L(0) \right)\\
&+r_\infty \left(\dfrac{n_-}{L}-z_L(0) \right)\left(\dfrac{n_+}{L}+z_L(0) \right)-\dfrac{1}{2}z_L(0)^2+\dfrac{\Delta_- I}{L^2}\,.
\end{aligned}
\end{equation}

The final result is best captured into the combinations $A_{iv_F}+A_{-iv_F}$ (appearing for even functions $\phi$) and $A_{iv_F}-A_{-iv_F}$ (for odd functions). Using the expression for $z_L(0)$ in eq.\eqref{zl0} we get
\begin{equation}
\label{result}
\begin{aligned}
A_{iv_F}+A_{-iv_F}&=\dfrac{-v_F}{12L^2}\dfrac{1}{\hat{s'}(iv_F)}\left(-1+3(n_++n_-)^2(1+2r_\infty)+3\dfrac{(n_+-n_-+2\varphi)^2}{1+2r_\infty} +12(\Delta_+I+\Delta_- I)  \right)\\
A_{iv_F}-A_{-iv_F}&=\dfrac{-v_F}{L^2}\dfrac{1}{\hat{s'}(iv_F)}\left(-\dfrac{1}{2}(n_++n_-)(n_+-n_-+2\varphi)+\Delta_-I-\Delta_+ I  \right)\,,
\end{aligned}
\end{equation}
at order $L^{-2}$, with $n_\pm$ given in eq.~\eqref{nnn} and $\Delta_\pm I$ in eq.~\eqref{iii}.

\subsection*{Double zero case\label{doublezero}}
We saw that if $1+\hat{r'}$ has a double zero at $\omega\in\Omega$ (meaning that moreover $(\hat{r'})'(\omega)=0$), then $\hat{w}_L$ can also include derivatives of the Dirac distribution and  eq.~\eqref{gen12} is modified. For the XXZ model this happens as soon as $\gamma/\pi$ is rational. But for most values of $\gamma$, the $\omega$ that are concerned are very large, and thus only the highest corrections in $L$ to the energy are modified. However for $\gamma=\pi/n$ with $n$ integer, the value $\pm iv_F$ is a double zero of $1+\hat{r'}$ and small changes have to be taken into account in the previous derivation. In this case we have
\begin{equation}
\hat{w}_L=-B_{iv_F}\delta'_{iv_F}-B_{-iv_F}\delta'_{-iv_F}+\sum_{\omega\in\Omega} A_\omega \delta_\omega\,,
\end{equation}
with $A_\omega$ and $B_{\pm iv_F}$ some constants. Since the Bethe root density \eqref{sigmaxxz} still has a simple pole at $\pm iv_F$, we need to have $\hat{s'}(\pm iv_F)=0$ and $(\hat{s'})'(\pm iv_F)\neq 0$. Then in the expression of the energy $e_L=-2\pi S_L(s')$ only the constants $B_{\pm iv_F}$ appear, not $A_{\pm iv_F}$. Back in eq.~\eqref{fcomplicated}, the Fourier transform of $f$ applied to $\omega\in\Omega$ selects the double poles in the negative half-plane of $\hat{\sigma}_\infty$ for $A_{\pm iv_F}$, and the simples poles in the negative half-plane of $\hat{\sigma}_\infty$ for $B_{\pm iv_F}$. Thus we get
\begin{equation}
w_L\left[(\delta_0+r')\star ((z_\infty-\alpha)\mathbbm{1}_{>z_L^{-1}(0)})\right]=\dfrac{1}{24 L^2}-\dfrac{(\hat{s'})'(-iv_F)}{v_F}B_{-iv_F}-\sum_{\omega\in\Omega_-}\dfrac{\hat{s'}(\omega)}{i\omega}A_\omega\,,
\end{equation} 
the sum over the $A_\omega$ being negligible at order $L^{-2}$ because of $\hat{s'}(-iv_F)=0$. Thus for eq.~\eqref{result} we get the same expression, but with $A_{\pm iv_F}$ replaced by $B_{\pm iv_F}$ and $\hat{s'}(iv_F)$ replaced by $(\hat{s'})'(\pm iv_F)$. Then the formula for $e_L$ hereafter \eqref{eL} is exactly the same.

\section{Corrections to observables}
The previous computations of $A_\omega$ can be used to determine the corrections to the energy, the momentum and the eigenvalue of the transfer matrix.
\subsubsection*{Energy}
Formula \eqref{result} gives the energy $e_L$ with $\phi=-2\pi s'$. We get:
\begin{equation}
\label{eL}
e_L=e_\infty-\dfrac{\pi v_F}{6L^2}\left(c-12(h+\bar{h}) \right)+o(L^{-2})\,,
\end{equation}
with
\begin{equation}
\begin{aligned}
c&=1-12\varphi^2 g^{-1}\\
h+\bar{h}&=\dfrac{1}{4}\left((n_++n_-)^2 g +(n_+-n_-+4\varphi)(n_+-n_-) g^{-1}  \right)+\Delta_+ I+\Delta_- I\\
g&=1+2r_\infty
\end{aligned}
\end{equation}
where we identify the central charge $c$ and the coupling constant $g=2(1-\gamma/\pi)$. $n_\pm$ are defined through eq.\eqref{nnn}.

 The $\Delta_\pm I$ terms deserve some comments. While the exponent with $\Delta_\pm I\neq 0$ naturally corresponds to descendants under the action of the Virasoro generators, eq.~\eqref{iii} should not conceal the fact that one does not get, in this way, all descendants in the corresponding Verma module. Indeed, taking the limits $\lambda\to\pm \infty$ in the counting function in eq.~\eqref{n0} for $\varphi=0$ shows that there is a bound on the possible Bethe numbers $I$
\begin{equation}
\label{bounddescendant}
|I|\leq \dfrac{L}{4}+(n_++n_-)\left(\dfrac{1}{2}-\dfrac{\gamma}{\pi} \right)\,.
\end{equation}
To see the consequences of this bound on the degeneracies of the descendants, let us discuss a simple example. Consider the case of an excited state corresponding to a primary field with $L/2-2$ Bethe roots, with $n_+=n_-=1$ for $\gamma=\pi/5$. There are two vacancies for positive Bethe roots and two for negative ones. Let us focus on the positive Bethe roots and denote by $\bullet$ a Bethe root, and by $+$ a vacancy. The right of the diagram corresponds to the highest Bethe number, and the left to the sea of Bethe roots. Then we have some descendants
\begin{equation}
\begin{aligned}
&\bullet\quad\bullet \quad \bullet\quad\bullet\quad \bullet\quad \bullet \quad \bullet \quad +\quad + \quad \quad \text{primary }\phi\\
&\bullet\quad\bullet \quad\bullet\quad\bullet\quad \bullet\quad \bullet \quad + \quad \bullet \quad + \quad \quad \text{descendant }L_{-1}\phi\\
&\bullet\quad\bullet \quad\bullet\quad\bullet\quad \bullet\quad \bullet \quad + \quad + \quad \bullet \quad \quad \text{descendant }L_{-2}\phi\\
&\bullet\quad\bullet \quad\bullet\quad\bullet\quad \bullet\quad + \quad \bullet \quad \bullet \quad + \quad \quad \text{descendant }L_{-1}^2\phi\\
&\bullet\quad\bullet \quad\bullet\quad\bullet\quad \bullet\quad + \quad \bullet \quad + \quad \bullet \quad \quad \text{descendant }L_{-2}L_{-1}\phi\\
&\bullet\quad\bullet \quad\bullet\quad\bullet\quad \bullet\quad + \quad + \quad \bullet \quad \bullet \quad \quad \text{descendant }L_{-2}^2\phi\\
&\bullet\quad\bullet \quad\bullet\quad\bullet\quad +\quad \bullet \quad + \quad \bullet \quad \bullet \quad \quad \text{descendant }L_{-2}^2L_{-1}\phi\\
\end{aligned}
\end{equation}
To identify the Bethe root configurations with Virasoro descendants we have used the following bijection:
The string of $L_{-n}$ operators, read from left to right, corresponds to the set of excited Bethe roots,
read from right to left, with $n$ being the number of vacancies seen to the left of a given Bethe root.
 In this example no $L_{-m}$ with $m\geq 3$ can be used, since there is no infinite sea (more precisely a number growing with $L$) of vacancies at the right. 
This constraint comes from the bound \eqref{bounddescendant}. Thus the number of different configurations of Bethe roots such that $\Delta^\pm I=k$ for a fixed $k$ depends on the number of positive vacancies $m$ and is $p_m(k)$ the number of ways (up to commutations) of writing $k$ as a sum of integers $\le m$. This integer $m$ is directly linked to the bound \eqref{bounddescendant}. For positive roots it is given by the integer part of $(n_++n_-)(1/2-\gamma/\pi)+n_++1/2$.

In the Virasoro algebra the number of descendants is independent of the sector, and is given by $p(k)$ the number of ways (up to commutations) of writing $k$ as a sum of positive integers. If there were an infinite sea of vacancies we would indeed recover the degeneracies of the Virasoro algebra. But here, the ``partial'' character $\chi_m(q)$ (partial because there may be other ways to get descendants) of the descendants that we can obtain with real Bethe roots for the sector with $m$ vacancies is then
\begin{equation}
 \chi_m(q)=\sum_{k=0}^\infty p_m(k)q^k=\prod_{k=1}^m\dfrac{1}{1-q^k}\,.
\end{equation}
In particular the ground state has no descendants with real Bethe roots, since in this case there is no vacancy at all. Notice as well that this partial character depends on $\gamma$ through $m$. We also stress that the fact that one does not recover the degeneracies of the Virasoro algebra with real Bethe roots has nothing to do with being in finite size $L$ instead of being in  the thermodynamic limit: even in this limit the bound \eqref{bounddescendant} remains valid. This is thus a priori not related to the finitized characters \cite{Finitized} which deal with the  conformal spectrum seen in finite size.

Some descendants that involves non-real Bethe roots have roots with imaginary part $i\pi/2$, and counting them is necessary to get the full degeneracies. But deriving the finite-size effects for such structures demands adaptations to the method that we do not treat in this paper.

\subsubsection*{Momentum}
We can apply the method to the momentum defined by $p_L=S_L(2i\pi s)$ as well. We get:
\begin{equation}
p_L=-\dfrac{2i\pi}{L} \alpha (n_+-n_-+2\varphi)+\dfrac{i\pi}{ L^2} (n_++n_-)(n_+-n_-+2\varphi)+\dfrac{2i\pi}{L^2}(\Delta_+ I-\Delta_- I)\,,
\end{equation}
that corresponds to
\begin{equation}
p_L=-\dfrac{2i\pi}{L} \alpha (n_+-n_-+2\varphi)+\dfrac{2i\pi}{ L^2} \left(h-\bar{h} \right)\,,
\end{equation}
with
\begin{equation}
h,\bar{h} = \dfrac{1}{8} \left((n_++n_-) \sqrt{g} \pm \dfrac{n_+-n_-+2\varphi}{\sqrt{g}} \right)^2 +\Delta_\pm I\,.
\end{equation}
Note that this allows us to identify the parameter $v_F$ with  the Fermi velocity, since it appears as the proportionality factor in $e\propto p$. Without $1/\sin\gamma$ in \eqref{H}, the Fermi velocity would be $\sin\gamma v_F$.

\subsubsection*{Eigenvalues}
The eigenvalue at spectral parameter $\mu$ of the transfer matrix for the corresponding six-vertex model is given by
\begin{equation}
\label{su2}
 \Lambda(\mu,\{\lambda_j\})=\sinh(\mu+i\gamma)^L\prod_{j=1}^{M}\dfrac{\sinh(\mu-\lambda_j-i\gamma/2)}{\sinh(\mu-\lambda_j+i\gamma/2)}+\sinh(\mu)^L\prod_{j=1}^{M}\dfrac{\sinh(\mu-\lambda_j+i3\gamma/2)}{\sinh(\mu-\lambda_j+i\gamma/2)}\,.
\end{equation}
Set $\mu=i\lambda$. For $-\gamma/2<\lambda<0$ the second term is exponentially smaller than the first term for the ground state and first excitations. Then the log of the absolute value of the eigenvalue is equal to the log of the first term up to exponentially small corrections. Denoting $f_L(\lambda)=\log(|\Lambda(i\lambda)|)/L$, we thus have
\begin{equation}
f_L(\lambda)=\log \sin(\lambda+\gamma) +S_L(F_\lambda)\,,
\end{equation}
with
\begin{equation}
F_\lambda(\mu)=\log \left| \dfrac{\sinh(i\lambda-\mu-i\gamma/2)}{\sinh(i\lambda-\mu+i\gamma/2)} \right|\,.
\end{equation}
The Fourier transform of this function can be evaluated \cite{Bateman}
\begin{equation}
\hat{F}_\lambda(\omega)=-\dfrac{\sinh(\lambda \omega)}{\omega}2\pi\hat{s'}(\omega)\,.
\end{equation}
 Then we have at order $L^{-2}$
\begin{equation}
f_L(\lambda)-f_\infty(\lambda)=\dfrac{\sin \lambda v_F}{v_F}(e_L-e_\infty)+o(L^{-2})\,.
\end{equation}
The spectral parameter plays the role of an anisotropy, and in the thermodynamic limit it amounts to rescaling one of the axes.

\section{Concluding remarks}
As a conclusion we give here an overview of the differences between our approach and the Wiener-Hopf and NLIE methods mentioned  in the introduction.
\subsection{Wiener-Hopf}
The starting point of this method is to use Euler-MacLaurin formula to express the sum of a function over the Bethe roots (which is $S_L(\phi)$ in our notations) \cite{WoynaEckle, DevegaWoyna, HamerQuispelBatchelor}. As already said, this operation is a Riemann sum of $\phi\circ z_L^{-1}$, and the Euler-MacLaurin formula we stated applies to functions that do not depend on $L$. There actually exists another version of Euler-MacLaurin with a remainder term that can be applied to $L$-dependent functions, which reads (with $\Lambda_L$ the largest Bethe root)
\begin{equation}
S_L(\phi)=\int_{-\Lambda_L}^{\Lambda_L} \phi z_L'+\dfrac{\phi(\Lambda_L)+\phi(-\Lambda_L)}{2L}+\dfrac{\phi'(\Lambda_L)-\phi'(-\Lambda_L)}{12L^2\sigma_\infty(\Lambda_L)}+\epsilon_L(\phi)\,.
\end{equation}
In general there is no guarantee that the remainder term $\epsilon_L(\phi)$ is negligible compared to the other ones, and actually this is precisely not the case in Bethe equations: for $r'$ or $s'$ the remainder term would be of order $O(L^{-2})$ as well. To go over this difficulty the following trick is used \cite{DevegaWoyna}: one adds $\hat{r'}\star z_L'$ to the equation defining $z_L'$ and then solve it for $z_L'$. This process creates the so-called 'dressed' functions $\phi^{\rm dr}$ given by
\begin{equation}
\hat{\phi^{\rm dr}}=\dfrac{\hat{\phi}}{1+\hat{r'}}\,.
\end{equation}
The Euler-MacLaurin fomula is then applied with these dressed functions. Although it was thought at the beginning that the remainder term is negligible after the dressing, it is actually still not the case, as pointed out by Karowski \cite{Karowski}. Nevertheless carrying out the computations without taking care of these terms still used to work, and Karowski gave some arguments to justify it. But according to \cite{Karowski} the intermediate steps are not numerically completely exact. Using the method of dressing an equation is obtained, involving $\Lambda_L$ and $z_L'(\Lambda_L)$ which are unknowns. To determine them a Wiener-Hopf equation is derived on $\chi(\lambda)=z_L'(\lambda+\Lambda_L)$
\begin{equation}
\chi(\lambda)+\int_0^{+\infty}\chi(s)k(\lambda-s)ds=f(\lambda)\,,
\end{equation}
with $k$ and $f$ some functions (that depend on $\Lambda_L$ and $z_L'(\Lambda_L)$). This is a very non-trivial equation to solve and demands complex analysis theorems \cite{WH2}. But solving it then leads to the result for the central charge after some work.\\

\subsection{NLIE}
Originally the NLIE were derived in \cite{NLIEKlumperBatchelor,PearceKlumper,Klumper} from analyticity properties of the eigenvalue and of some auxiliary functions. A shortcut was then found \cite{DestriDevega} by expressing $S_L(\phi)$ as a contour integral thanks to the residue theorem
\begin{equation}
S_L(\phi)=\dfrac{1}{2i\pi L}\oint_{\mathcal{C}} \phi(x)\dfrac{d}{dx}\log\left(1+e^{2i\pi Lz_L(x)} \right)dx\,,
\end{equation}
where $\mathcal{C}$ is a contour that encircles the Bethe roots. The integrand indeed has a pole at each Bethe root, since by definition $e^{2i\pi Lz_L(x)}$ evaluates to $-1$ at them (when Bethe numbers are half-integers). Then by expressing $S_L(r(\lambda-\cdot))$ this way, a non-linear integral equation can be found for $z_L$.

 The energy is then expressed in a similar way. As in the Wiener-Hopf case, a manipulation is made so that to involve the dressed functions $\phi^{\rm dr}$. The result (in particular the central charge $c=1$) is then obtained through dilogarithm identities.

\subsection{Differences}
We expand here the comment made in the introduction on the differences between these methods and the one we present.\\

Our method mainly uses the tools of distribution theory. No Wiener-Hopf equations, no dilogarithms and very little complex analysis are used. As it is common with this technology, we start by studying the functional $S_L$ when applied on $\mathcal{C}^\infty$ functions with compact support. A powerful constraint given by equation \eqref{gen12} is derived. Then it is extended to more general functions with possible discontinuities and without compact support. This observation is at the heart of our approach and is totally absent from the two previous methods. \\

To compute the main constants $A_{\pm iv_F}$ that appear in \eqref{gen12}, we sum the counting functions over positive or negative Bethe roots. No equivalent operation can be found in the previous methods as well, which both focus on the properties of the tail of the counting function. Note that beside getting the central charge and conformal dimensions in an efficient way this operation permits us to get the finite-size effects for the descendants very simply.\\

Possible directions for applications or extensions  of the method include treating the  case of complex roots (strings), treating configurations with isolated Bethe roots, or computing next-order corrections, including logarithmic ones. Complex roots that appear in some higher-rank or higher-spin systems demand a suitable adaptation of the Euler-MacLaurin formula, which is no longer applicable in this case. Configurations with isolated Bethe roots are relevant in particular for the analysis of Verma modules, since they are necessary to describe  descendants in the XXZ model. Finally it can be seen that some of the next-order corrections are quadratic in the $A_\omega$'s, and in case of a non-invertible kernel $1+2r_\infty$ the fact that $0\in\Omega$ in eq.\eqref{omega} together with the bound \eqref{bound} leaves room for logarithmic corrections. We hope that progress on these various directions will be discussed in subsequent work.\\

\noindent{\bf Acknowledgments:} this work was supported in part by the ERC Advanced Grant NuQFT. We thank T. Silva, Y. Ikhlef and C. Kopper for discussions. 

\bibliographystyle{hplain}
\bibliography{finite_size_bethe}

\end{document}